\documentclass[a4paper,11pt]{article}
\usepackage{amsfonts}
\usepackage{amsmath}
\usepackage{amssymb}
\usepackage{graphicx}
\usepackage{array}
\usepackage{booktabs}
\usepackage{multirow}
\usepackage{makecell}
\usepackage{float}
\pdfoutput=1 % if your are submitting a pdflatex (i.e. if you have
\usepackage{mymacros}
\usepackage{chngcntr}
\usepackage{verbatim}
\usepackage{amsthm}
\usepackage{graphics}
\usepackage{xcolor}
\usepackage{mathrsfs}
\usepackage{bbold}
\usepackage{cases}
\usepackage{epsfig}
\usepackage{epstopdf}
\usepackage{hyperref}
\usepackage{amsfonts}
\usepackage{hyperref}
\usepackage{jheppub} % for details on the use of the package, please
\usepackage[T1]{fontenc} % if needed

\hypersetup{
	colorlinks=true,
	linkcolor=blue,
	filecolor=blue,
	urlcolor=blue,
	citecolor=blue,
}

\title{\boldmath Lyapunov exponents, phase transition and chaos bound in Kerr-Newman AdS spacetime}

\author{Chuang Yang$^{a}$,\footnote{chuangyangyc@hotmail.com}}
\author{Chuanhong Gao$^{b}$,\footnote{chuanhonggao@hotmail.com}}
\author{Deyou Chen$^{a}$,\footnote{deyouchen@hotmail.com}}
\author{Xiaoxiong Zeng$^{c}$\footnote{xxzengphysics@163.com}}

\affiliation{$^{a}$School of Science, Xihua University, Chengdu 610039, China}
\affiliation{$^{b}$Faculty of Engineering and Quantity Surveying, INTI International University, Persiaran Perdana BBN, Putra Nilai, Nilai 71800, Negeri Sembilan, Malaysia}
\affiliation{$^{c}$College of Physics and Electronic Engineering, Chongqing Normal University, Chongqing 401331, China}

\abstract{In this paper, we investigate Lyapunov exponents associated with chaotic motions of both massless and massive particles in the vicinity of a Kerr-Newman AdS black hole. Our exploration focuses on their correlations with the black hole phase transition and the chaos bound. The results demonstrate that these exponents serve as effective probes of the phase transition, with the chaotic Lyapunov exponent of the massless particle offering a more precise characterization. Further calculations indicate that critical exponents linked to these Lyapunov exponents are uniformly 1/2. Notably, the violation of the chaos bound occurs irrespective of whether a phase transition is taking place. Through comparative analysis, we identify a critical radius, and the violation consistently arises when the black hole's radius is less than this critical radius. Moreover, this violation is observed in the spacetime of the stable small black hole during the phase transition.}

\begin{document} 
	\maketitle
	\flushbottom
	
\section{Introduction}

Phase transition has always been an important focus in the realm of thermodynamics researches. The exploration of black hole(BH) phase transitions originates from the Hawking-Page phase transition, which revealed a first-order phase transition occurring between the Schwarzschild BH and the thermal anti-de Sitter(AdS) vacuum in AdS space \cite{SWHP}. When the temperature drops below the critical value, the thermal AdS vacuum exhibits greater stability compared to the BH. Conversely, when the temperature surpasses the critical threshold, the BH attains a higher level of stability than the thermal AdS vacuum. Subsequently, taking into account a varying cosmological constant\cite{KRT1,KRT2,KRT3}, people delved into the thermodynamic behaviors of a series of BHs in the extended phase space \cite{CCLY,WL1,WL2}. By establishing the equation of state for BHs, they studied the behavior of the phase transitions and obtained the critical exponents when the phase transitions occur. These critical exponents were found to exactly match those of the classical van der Waals fluid. Meanwhile, certain BHs exhibit triple points and undergo triple-point phase transitions \cite{RGC,WLWL,FKM1,FKM2}, accompanied by oscillatory phenomena during the phase transition process. Reentrant phase transitions were also found in some BHs \cite{AKM1,ZYZ}. For other meaningful thermodynamic phenomena, please refer to References \cite{XCH1,XCH2,XCH3,XCH4,XCH5,XCH6,XCH7,XCH8,XCH9,XCH10,XCH11,XCH12,XCH13,XCH14,XCH15,XCH16,CLL1,CLL2,CLL3,CLL4,CLL5,CLL6,CLL7,CLL8,CLL9,CLL10,CLL11}, as well as the references cited within these articles.

A Lyapunov exponent(LE) serves as a direct measure of a system's sensitivity to initial conditions by quantifying the temporal evolution of the distance between adjacent trajectories. It stands as an important indicator for assessing the chaotic sensitivity of the system.  A positive LE signifies that the separation between adjacent trajectories increases exponentially over time, indicative of chaotic behavior within the system. Conversely, a negative LE implies an exponential decay in the distance between adjacent trajectories, corresponding to stable motion of the system. The exponent finds applications not only in the identification of chaotic behavior in BH spacetimes but also plays a pivotal role in other domains, including the quantization of BHs' areas and entropies, as well as the determination of quasi-normal modes.

Recently, through an investigation into the correlation between the LEs of particles' and string's chaos in the vicinity of the Reissner-Nordstrom AdS BH and the phase transition of the BH \cite{GLMW}, the scope of the exponent's utility has been significantly broadened. The investigation reveals that the multi-valued nature of the exponent can effectively mirror the phase-transition behavior. In the absence of a phase transition, the exponent retains a single-valued characteristic. Furthermore, the discontinuity of the exponent has been identified as an order parameter characterized by a critical exponent of 1/2. Given that this BH shares thermodynamic properties akin to those of a van der Waals fluid, the exponent offers a dynamical explanation for this resemblance. Subsequently, this research has been extended to other spherically symmetric spacetimes \cite{GLMW1,GLMW2,GLMW3,GLMW4,GLMW5,GLMW6}, further elucidating the probing capability of the exponent in the context of BH phase transitions.

In this paper, we investigate the LEs for chaos exhibited by both massless and massive particles in the vicinity of a Kerr-Newman AdS BH, to explore the relationship between these exponents and the phase transition of this BH. To further unravel their connections, we compute critical exponents associated with the LEs. Furthermore, we study the upper bound of the LEs specifically during the phase transition. In the framework of the AdS/CFT correspondence, the thermodynamic properties of the Kerr-Newman AdS BH correspond to physical observables in the conformal field theory(CFT). The behavior of the LE in this spacetime may map onto the chaotic properties within CFT. Exploring this relationship facilitates the validation of the holographic duality theory and elucidates the manifestations of phase transitions in the context of CFT. The exploration into the upper bound of the LE can be traced back to \cite{MSS}. Within this innovative work, Maldacena, Shenker and Stanford formulated a conjecture. They posited that in thermal quantum systems endowed with a large number of degrees of freedom, there exists a universal upper bound, known as the chaos bound, on the exponent. Moreover, this upper bound is linked to the system's temperature.  Subsequently, Hashimoto and Tanahashi studied a single particle system in the vicinity of a BH \cite{HT}. Their findings demonstrated that the exponent does not exceed the BH's surface gravity. When considering the relationship between the surface gravity and temperature, this result aligns  with the conjecture proposed in \cite{MSS}. In subsequent research endeavors, people have incorporated the impacts of the charges and angular momenta of the particles and BHs into their investigations of the exponent. Their findings suggest the violation of the chaos bound \cite{ZLL,LGR1,LGR2,KG1,KG2,GCYW}.

The remainder of this paper is organized as follows. We investigate the thermodynamic of the Kerr-Newman AdS BH in the next section. In Section \ref{sec3}, the relationship between the LEs of the chaos for both massless and massive particles and the phase transition is explored, and the critical exponents associated with the LEs are calculated. In Section \ref{sec4}, we study the upper bound of the LE to examine whether the chaos bound is violated. The last section is devoted to our conclusions and discussions.

\section{Thermodynamics of Kerr-Newman AdS BH}\label{sec2}

The Kerr-Newman AdS BH describes a rotating and charged AdS spacetime, and its metric is given by 

\begin{eqnarray}
ds^2 =-\frac{\Delta}{\rho^2}\left(dt-\frac{a\sin^2\theta}{\Xi} d\varphi\right)^2 +\frac{\rho^2}{\Delta}dr^2 + \frac{\rho^2}{\Sigma} d\theta^2  +\frac{\Sigma\sin^2\theta}{\rho^2}\left(adt -\frac{r^2+a^2}{\Xi}d\varphi\right)^2,
\label{eq2.2.1}
\end{eqnarray}

\noindent with an electromagnetic potential

\begin{eqnarray}
A_{\mu}= A_tdt+A_{\varphi}d\varphi=\frac{Qr}{\rho^2}dt  -\frac{Qra\sin^2\theta}{\rho^2\Xi}d\varphi,
\label{eq2.2.2}
\end{eqnarray}

\noindent where

\begin{eqnarray}
\Delta &=& \left(r^2+a^2\right)\left(1+\frac{r^2}{l^2}\right) -2mr+Q^2,  \nonumber\\
\rho^2 &=& r^2 + a^2\cos^2\theta, \quad \Xi= 1-\frac{a^2}{l^2}, \quad \Sigma= 1-\frac{a^2}{l^2}\cos^2\theta.
\label{eq2.2.3}
\end{eqnarray}

\noindent $m$, $a$ and $Q$ are the BH's mass, rotational parameter and charge, respectively.  $l$ is the AdS radius and connects to the cosmological constant as $\Lambda=-\frac{3}{l^2}$  The ADM mass is $M=\frac{m}{\Xi^2}$, and the angular momentum is $J=\frac{ma}{\Xi^2}$.  The entropy, Hawking temperature and angular velocity are

\begin{eqnarray}
S &=& \frac{\pi(r_+^2 + a^2)}{\Xi},\\
T &=& \frac{r_+(1+ \frac{a^2}{l^2}+\frac{3r_+^2}{l^2}-\frac{a^2+Q^2}{r_+^2})}{4\pi (r_+^2+a^2)}, \label{eq2.2.5}\\
\Omega &=& \frac{a(l^2+r_+^2)}{l^2(r_+^2+a^2)},
\label{eq2.2.6}
\end{eqnarray}

\noindent respectively, where $r_+$ is the event horizon determined by the largest positive root of $\Delta=0$. The unfixed cosmological constant can reconcile the inconsistency between the first law of thermodynamics of BHs and the Smarr relation derived from the scaling method. When this constant is treated as the thermodynamic pressure $P = \frac{3}{8\pi l^2}$ in the Kerr-Newman AdS spacetime, its conjugate quantity is the thermodynamic volume 

\begin{eqnarray}
V= \frac{2\pi \left[(r_+^2+a^2)(2r_+^2 l^2+a^2 l^2-r_+^2 a^2)+l^2 Q^2 a^2\right]}{3l^2 \Xi^2 r_+}.
\label{eq2.2.7}
\end{eqnarray}

\noindent The above thermodynamic quantities obey the first law of thermodynamics
\begin{eqnarray}
dM = TdS+ \Omega dJ+ \Phi dQ+ VdP,
\label{eq2.2.8}
\end{eqnarray}\noindent 

\noindent and the corresponding Smarr relation is given by

\begin{eqnarray}
M = 2TS + 2\Omega J+ \Phi Q -2 VP.
\label{eq2.2.9}
\end{eqnarray}

\noindent The Gibbs free energy for this BH is

\begin{eqnarray}
F &=& M-TS 
\nonumber\\ &=& \frac{-l^2r_+^4+a^4(-l^2+r_+^2)+l^4(3q^2+r_+^2)+a^2\left[3l^4+3r_+^4-l^2(Q^2-2r_+^2)\right]}{4(a^2-l^2)^2r_+}.
\label{eq2.2.10}
\end{eqnarray}

From dimensional analysis, we find that the scaling of these physical quantities is a power of $l$,

\begin{eqnarray}
r =\bar{r}l, \ r_+ =\bar{r}_+l, \ a= \bar{a}l, \ Q=\bar{Q}l, \ m= \bar{m}l, \ J= \bar{J}l^2, \ T = \bar{T}l^{-1}, \ F= \bar{F}l,
\label{eq2.2.11}
\end{eqnarray}

\noindent The above physical quantities with bars represent dimensionless quantities. Eq. (\ref{eq2.2.6}) demonstrates the relation between the temperature and horizon radius. When a single temperature corresponds to multiple horizon radii, it suggests that the BH can manifest in multiple phases at that particular temperature. On the contrary, if a temperature is linked to just one radius, it signifies that the BH is in a single phase at that temperature. To study the phase transition, we fix the angular momentum at $\bar{J}=0.01$ and first find the critical point which is determined by

\begin{eqnarray}
\left(\frac{\partial \bar{T}}{\partial \bar{r}_+}\right)_{\bar{Q}}=\left(\frac{\partial^2 \bar{T}}{\partial \bar{r}_+^2}\right)_{\bar{Q}}=0.
\label{eq2.2.12}
\end{eqnarray}

\noindent At the critical point, the relevant physical quantities assume the following values

\begin{eqnarray}
\bar{r}_{+c} = 0.368117, \quad \bar{Q}_c=0.146968, \quad \bar{T}_c=0.263909.
\end{eqnarray}

\begin{figure}[h]
	\centering
	\includegraphics[width=10cm,height=7cm]{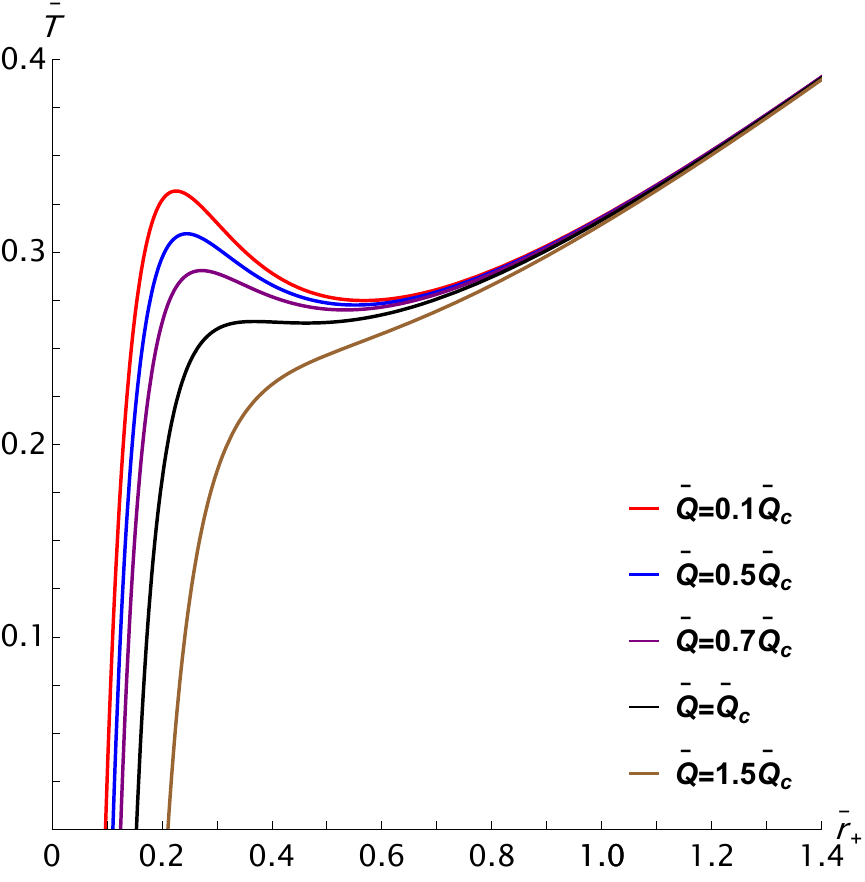}
	\caption{The temperature of the Kerr-Newman AdS BH varies with the horizon radius, where $\bar{J}=0.01$.}
	\label{4f1}
\end{figure}

By employing Eqs. (\ref{eq2.2.6}) and (\ref{eq2.2.11}), we plot the curve of the temperature variation with respect to the radius, as illustrated in Figure \ref{4f1}. From the figure, we can see that when $\bar{Q}<\bar{Q}_c$, the temperature curve displays a maximum and a minimum. This suggests that the BH may exist in multiple phases, and phase transitions are likely to occur for these charge values. When  $\bar{Q}>\bar{Q}_c$, the temperature curve is a monotonically increasing curve with respect to the horizon radius, implying that no phase transition occurs.
 
 \begin{figure}[h]
 	\centering
 	\begin{minipage}[t]{0.48\textwidth}
 		\centering
 		\includegraphics[width=7cm,height=5cm]{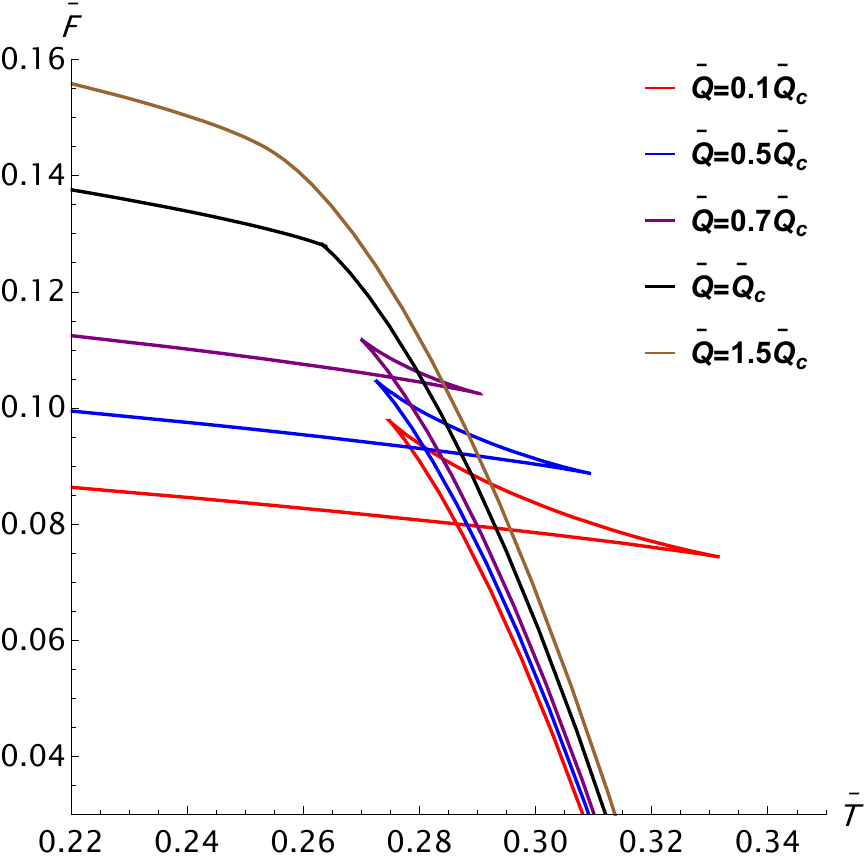}
 		\subcaption{}
 		\label{4-2-a}
 	\end{minipage}
 	\begin{minipage}[t]{0.48\textwidth}
 		\centering
 		\includegraphics[width=7cm,height=5cm]{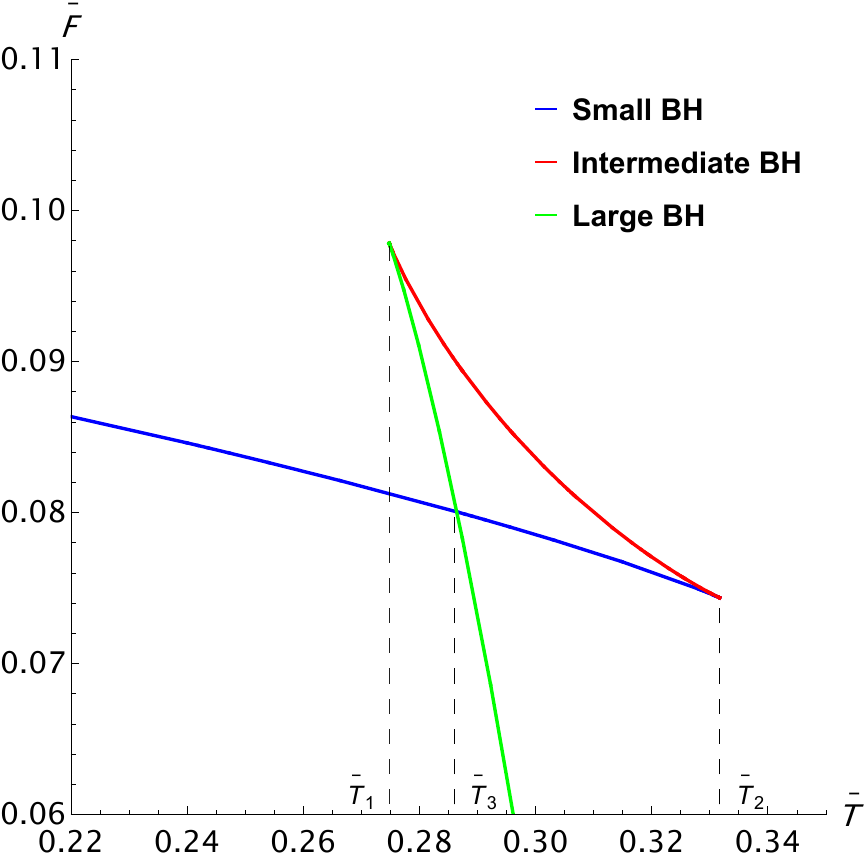}
 		\subcaption{}
 		\label{4-2-b}
 	\end{minipage}
 	\caption{The free energy of the Kerr-Newman AdS BH varies with the temperature, where $\bar{J}=0.01$. Figure \ref{4-2-b} shows the case where $\bar{Q}=0.1\bar{Q}_c$.}
 	\label{4f2}
 \end{figure}
 
To investigate the phase transition, we plot the Gibbs free energy as a function of the temperature for different values of the charge in the canonical ensemble of the Kerr-Newman AdS BH in Figure \ref{4f2}. As shown in Figure \ref{4-2-a}, when $\bar{Q}<\bar{Q}_c$, a swallowtail structure emerges, indicating an occurrence of a phase transition in the BH. Conversely, when $\bar{Q}>\bar{Q}_c$, the swallowtail structure disappears and there is no phase transition. Figure \ref{4-2-b} illustrates the relationship between the free energy and the temperature when $\bar{Q}=0.1\bar{Q}_c$. The graph clearly shows three distinct branches corresponding to small, intermediate, and large BHs.
When $\bar{T}<\bar{T}_1$ and $\bar{T}>\bar{T}_2$, the free energy is single valued. However, it becomes multi-valued within the intermediate temperature range $\bar{T}_1<\bar{T}<\bar{T}_2$. 
The phases of the small BH and the large BH are stable, while the phase of the intermediate BH is unstable. As the temperature varies, the first-order phase transition between the small and large BH phases occurs at $\bar{T}_3$.

\section{LEs and phase transition of Kerr-Newman AdS BH}\label{sec3}

LEs are crucial concepts in chaos theory, serving to quantify the rate at which trajectories in dynamical systems diverge or converge. In this section, we study the LEs associated with the chaotic motions of massless and massive particles within the equatorial plane of the Kerr-Newman AdS BH and their relationships with thermodynamic phase transitions.

\subsection{LEs}\label{sec3.1}

We begin by deriving the Lagrangian that describes the motion of a neutral particle in the equatorial plane of the BH. The derived Lagrangian can be expressed in the following form

\begin{eqnarray}
	\mathcal{L} = \frac{1}{2}g_{\mu\nu} \dot{x}^{\mu}\dot{x}^{\nu}= \frac{1}{2}\left(\bar{g}_{tt}\dot{t}^2+\bar{g}_{rr}\dot{r}^2 +\bar{g}_{\varphi\varphi}\dot{\varphi}^2+2\bar{g}_{t\varphi}\dot{t}\dot{\varphi}\right).
	\label{eq2.2.13}
\end{eqnarray}

\noindent where the dot denotes the derivative with respect to proper time, $\bar{g}_{\mu\nu}$ represent the metric functions on the equatorial plane and are given by

\begin{eqnarray}
	\bar{g}_{tt} = -\frac{\Delta - a^2}{r^2},\quad
	\bar{g}_{rr} =  \frac{r^2}{\Delta},\quad
	\bar{g}_{\varphi\varphi}  = \frac{(r^2+a^2)^2 -\Delta a^2}{r^2\Xi^2},\quad
	\bar{g}_{t\varphi} = -\frac{(r^2+a^2-\Delta)a}{r^2\Xi}.
	\label{eq2.2.14}
\end{eqnarray}

\noindent From the Lagrangian, we readily compute canonical momenta of the particle as 

\begin{eqnarray}
	p_t &=& \frac{\partial\mathcal{L}}{\partial\dot{t}} = \bar{g}_{tt}\dot{t}+ \bar{g}_{t\varphi}\dot{\varphi}=-E, \label{eq:2.2.15} \\
	p_r &=& \frac{\partial\mathcal{L}}{\partial\dot{r}} = \bar{g}_{rr}\dot{r}, \label{eq:2.2.16} \\
	p_{\varphi} &=& \frac{\partial\mathcal{L}}{\partial\dot{\varphi}} = \bar{g}_{\varphi\varphi}\dot{\varphi}+ \bar{g}_{t\varphi}\dot{t} =L, \label{eq:2.2.17}
\end{eqnarray}

\noindent where $E$ and $L$ are the energy and angular momentum of the particle. From Eqs. \eqref{eq:2.2.15} and \eqref{eq:2.2.17}, we derive the specific expressions for $\dot{t}$ and $\dot{\varphi}$, which are

\begin{eqnarray}
	\dot{t} &=& \frac{E \bar{g}_{\varphi\varphi}+L\bar{g}_{t\varphi}}{\bar{g}_{t\varphi}^2-\bar{g}_{tt}\bar{g}_{\varphi\varphi}}, \\
	\dot{\varphi} &=& \frac{E\bar{g}_{t\varphi}+L\bar{g}_{tt}}{\bar{g}_{tt}g_{\varphi\varphi}-\bar{g}_{t\varphi}^2}.
	\label{eq2.2.18}
\end{eqnarray}

\noindent The Hamiltonian is

\begin{eqnarray}
	2\mathcal{H} = 2(p_{\mu}\dot{x}^{\mu}-\mathcal{L}) = p_{\mu}\dot{x}^{\mu}= \bar{g}_{tt}\dot{t}^2+\bar{g}_{rr}\dot{r}^2 +\bar{g}_{\varphi\varphi}\dot{\varphi}^2+2\bar{g}_{t\varphi}\dot{t}\dot{\varphi}=\delta,
	\label{eq2.2.19}
\end{eqnarray}

\noindent where $\delta = 0$ corresponds to a null geodesic, and $\delta = -1$ to a timelike geodesic. Solving Eq. (\ref{eq2.2.19}), we get $r$-motion as follows

\begin{eqnarray}
	\dot{r}^2 = \frac{\delta}{\bar{g}_{rr}}-\frac{E^2 \bar{g}_{\varphi\varphi} +L^2 \bar{g}_{tt} +2EL\bar{g}_{t\varphi} }{\bar{g}_{rr}(\bar{g}_{tt}\bar{g}_{\varphi\varphi}-\bar{g}_{t\varphi}^2)}.
	\label{eq2.2.20}
\end{eqnarray}

We adopt the method in \cite{LE1,LE2,LE3,ZLL,LGR1} to derive the LE. We first define an effective potential as $V_r=\dot{r}^2$, and then derive the LE in the following form from this effective potential,

\begin{eqnarray}
\lambda_c = \sqrt{\frac{ V^{\prime\prime}_r}{2\dot{t}^2}}.
\label{eq2.2.21}
\end{eqnarray}

\noindent where the prime denotes the derivative with respect to $r$. This LE reflects the instability of the particle motion in the equatorial planes of BHs.

\subsection{Null geodesic's case}\label{sec4.1}

For a massless particle, we utilize the definition of the effective potential  $V_r = \dot{r}^2$ and $\delta = 0$ to derive the potential as

\begin{eqnarray}
	V_{r}= \frac{(aE-L\Xi)^2\Delta-\left[E(r^2+a^2)-aL\Xi_a)\right]^2}{r^4}.
	\label{eq2.3.1}
\end{eqnarray}

\begin{figure}[h]
	\centering
	\begin{minipage}[t]{0.48\textwidth}
		\centering
		\includegraphics[width=7cm,height=5cm]{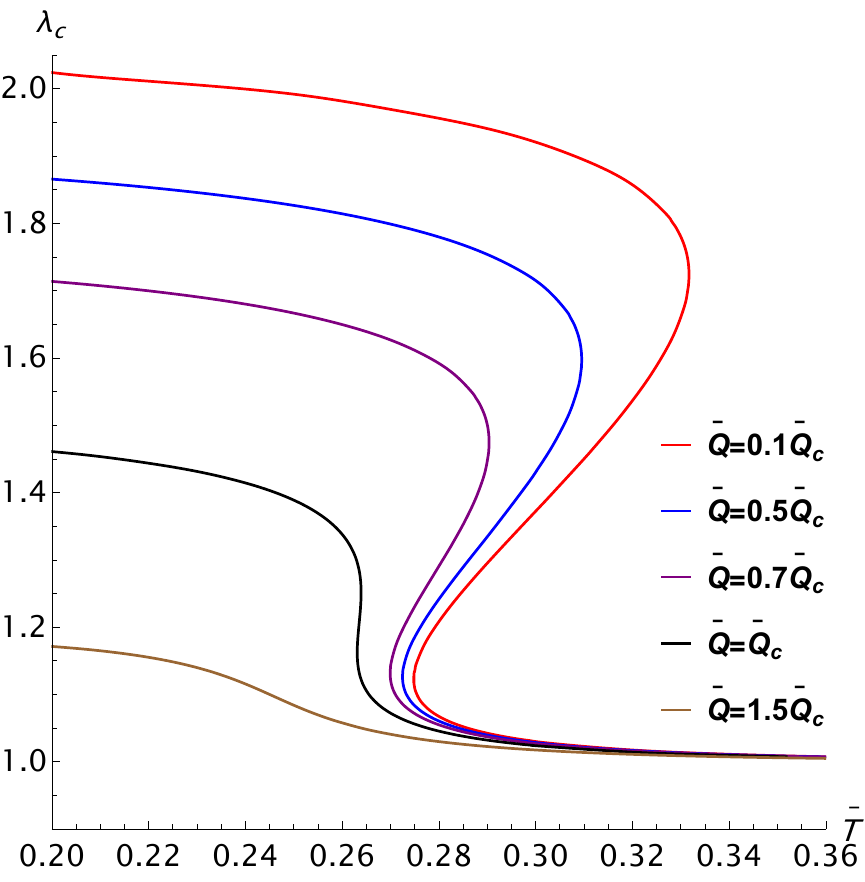}
		\subcaption{}
		\label{4-4-a}
	\end{minipage}
	\begin{minipage}[t]{0.48\textwidth}
		\centering
		\includegraphics[width=7cm,height=5cm]{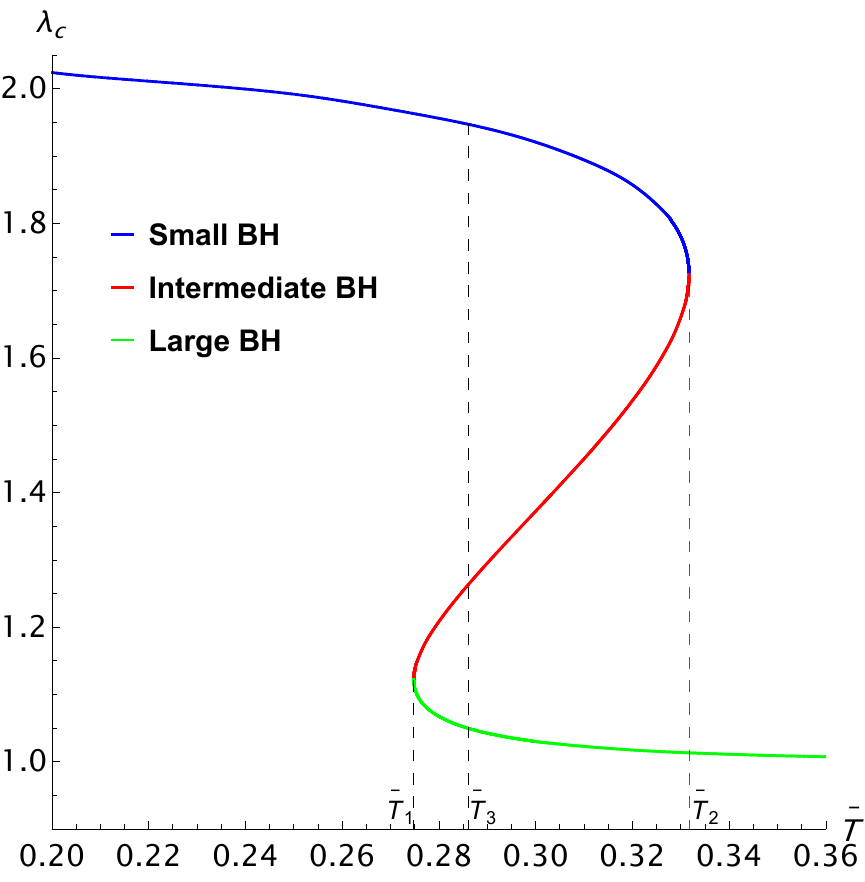}
		\subcaption{}
		\label{4-4-b}
	\end{minipage}
	\caption{The LEs of chaos for the massless particle around the Kerr-Newman AdS BH varies with the temperature $\bar{T}$. Figure \ref{4-4-b} shows the case where $\bar{Q}=0.1\bar{Q}_c$. }
	\label{4f4}
\end{figure}

\noindent Chaos of particles emerges from their unstable equilibrium orbits, therefore, we first identify the locations of these orbits, which satisfy $V_{r}=V_{r}^{\prime}=0$ and $V_{r}^{\prime\prime}>0$. From Eqs. (\ref{eq2.2.14}), (\ref{eq2.2.21}) and (\ref{eq2.3.1}), we obtain the LE, and then generate curves to depict its variation with respect to the temperature in Figures \ref{4f4}. For the calculations presented in this paper, we order $L=20l$ and $\bar{J}=0.01$.

Figure \ref{4f4} depicts how the LE varies with the temperature under different charge values $\bar{Q}$. In Figure \ref{4-4-a}, the exponent is multi-valued only when $\bar{Q} < \bar{Q}_c$, and it becomes single-valued and decreases monotonically with temperature when $\bar{Q} > \bar{Q}_c$. Figure \ref{4-4-b} specifically illustrates the variation of the exponent with the temperature when  $\bar{Q} = 0.1\bar{Q}_c$. The same color coding as in Figure \ref{4f2} has been adopted here: the blue curve represents small BHs, the red curve signifies intermediate-sized BHs, and the green curve denotes large BHs. From the figure, it is observed that when the temperature $\bar{T}$ increases from 0 to $\bar{T}_1$, the exponent is single-valued and gradually decreases, with the small BH phase in this temperature range. In the range of $\bar{T}_1<\bar{T}<\bar{T}_2$, there are three distinct values of $\lambda$ for a temperature $\bar{T}$, indicating the coexistence of small BH, intermediate BH, and large BH phases. This coexistence implies that these phases can mutually transform within this temperature range. When $\bar{T}>\bar{T}_2$, the exponent becomes single-valued again and monotonically decreases, with the large BH phase. The temperature $\bar{T}_3$ marks a first-order phase transition between the small and large BHs. These behaviors of the exponent in Figure \ref{4f4} closely resemble the free energy's behavior observed in Figure \ref{4f2}, demonstrating that the LE can effectively probe the phase transitions of the Kerr-Newman AdS BH.

\subsection{Timelike geodesic's case }\label{sec3.2}

For a massive particle, we get $\delta=-1$ in Eq. (\ref{eq2.2.20}). Then the effective potential for this particle in the Kerr-Newman AdS spacetime is

\begin{eqnarray}
	V_{r} = \frac{-\Delta r^2+(aE-L\Xi)^2\Delta-\left[E(r^2+a^2)-aL\Xi_a)\right]^2}{r^4}.
	\label{eq2.3.2}
\end{eqnarray}

\begin{figure}[h]
	\centering
	\begin{minipage}[t]{0.48\textwidth}
		\centering
		\includegraphics[width=7cm,height=5cm]{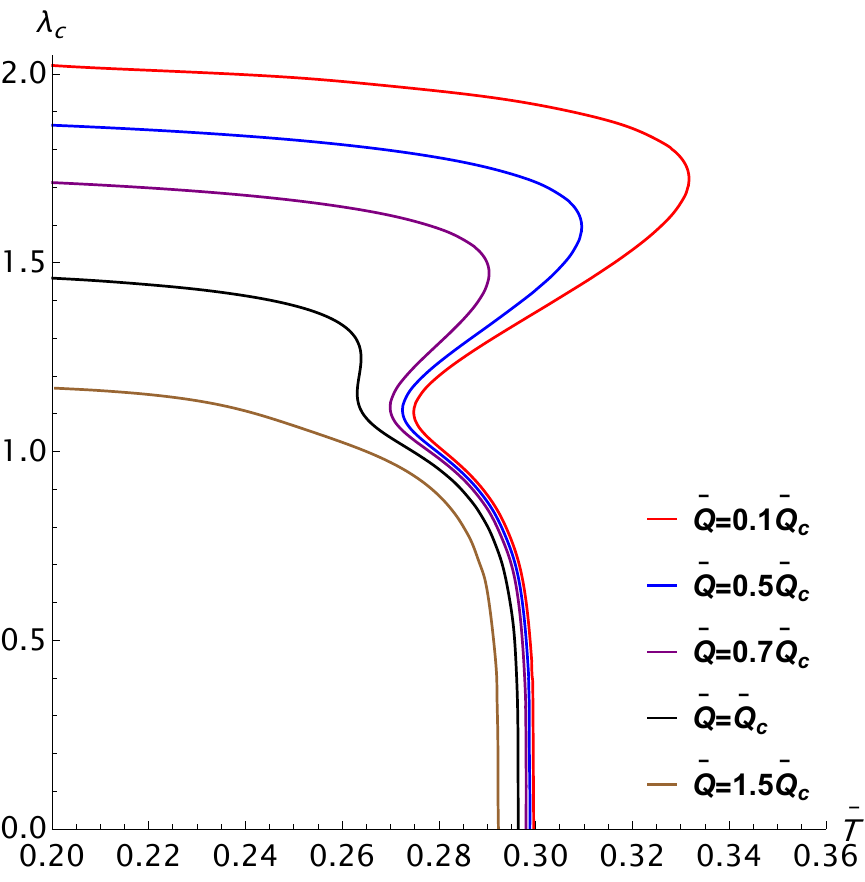}
		\subcaption{}
		\label{4-6-a}
	\end{minipage}
	\begin{minipage}[t]{0.48\textwidth}
		\centering
		\includegraphics[width=7cm,height=5cm]{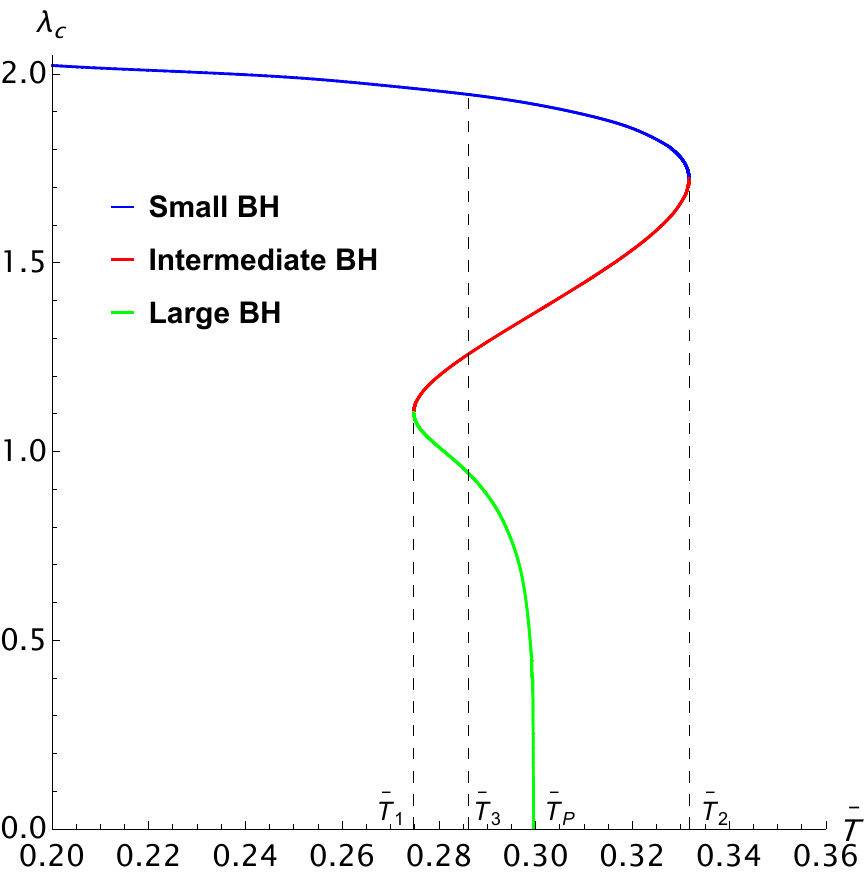}
		\subcaption{}
		\label{4-6-b}
	\end{minipage}
	\caption{The LEs of chaos for the massive particle around the Kerr-Newman AdS BH varies with the temperature $\bar{T}$. Figure \ref{4-6-b} shows the case where $\bar{Q}=0.1\bar{Q}_c$.}
	\label{4f6}
\end{figure}

\noindent We use Eqs. (\ref{eq2.2.14}), (\ref{eq2.2.21}) and (\ref{eq2.3.2}) to derive the LE of chaos for the massive particle. The curves to depict its variation with respect to the temperature in Figure \ref{4f6}.

We use Eqs. (\ref{eq2.2.14}), (\ref{eq2.2.21}) and (\ref{eq2.3.2}) to derive the LE of chaos for the massive particle. Figure \ref{4-6-a} depicts the variation of the exponent with the temperature for different charge values $\bar{Q}$. Similarly, we observe that the exponent is a multi-valued function when $\bar{Q}<\bar{Q}_c$. It becomes single-valued when $\bar{Q}>\bar{Q}_c$, indicating the presence of a phase transition. Figure \ref{4-6-b} demonstrates the relationship between the exponent and temperature for the specific case of $\bar{Q}=0.1\bar{Q}_c$. When $\bar{T}<\bar{T}_1$, the exponent monotonically decreases with increasing the temperature. In the range $\bar{T}_1<\bar{T}<\bar{T}_2$, the exponent exhibits a multi-valued behavior. Unlike the massless particle's case, the LE for the massive particle's chaos features a termination temperature $\bar{T}_{p}$. At $\bar{T}=\bar{T}_{p}$, the exponent equals zero, and the unstable equilibrium orbit vanishes. Within the range $\bar{T}_1<\bar{T}<\bar{T}_{p}$, three distinct values of the exponent exist for the same temperature, corresponding to the small, intermediate, and large BH phases. These phases coexist and can be transformed into each other. In the range $\bar{T}_{p}<\bar{T}<\bar{T}_2$, two different values are observed per temperature, representing the small and intermediate BH phases. Due to the appearance of the termination temperature $\bar{T}_{p}$, the large BH phase (specifically the portion with $\bar{T}>\bar{T}_{p}$) cannot be fully described by the exponent.

\subsection{Critical exponents}\label{sec3.2.3}
In this section, we investigate the critical exponents of the phase transition for the Kerr-Newman AdS BH. Here, we employ the concise and elegant method proposed in \cite{BR1,BR2} to calculate the critical exponents associated with the LEs for chaos of the massless and massive particles. Throughout this section, the subscript 'c' denotes the values evaluated at the critical point.

We try to find critical exponents $\alpha_1$ and $\alpha_2$ that satisfy $\Delta\lambda$, 
\begin{eqnarray}
	\Delta\lambda \sim \left |\bar{Q}-\bar{Q}_i \right |^{\alpha_1}, \quad
	\Delta\lambda \sim \left |\bar{T}-\bar{T}_i \right |^{\alpha_2}.
	\label{eq2.4.1}
\end{eqnarray}

\noindent Here $\lambda$ denotes the LEs of chaos for the massless and massive particles. The horizon radius $\bar{r}_+$ at the phase transition point can be expressed in terms of the horizon radius $\bar{r}_i$ at the critical point as,

\begin{eqnarray}
	\bar{r}_+=\bar{r}_i\left (1+\sigma \right)
	\label{eq2.4.2}
\end{eqnarray}

\noindent Where $\left |\sigma \ll1\right|$. The charge $\bar{Q}$ of the BH can be expressed as a function of the horizon radius $\bar{r}_+$, denoted as $\bar{Q}\left ( \bar{r}_+ \right)$. Similarly, the charge $\bar{Q}\left ( \bar{r}_+ \right)$ can be written as

\begin{eqnarray}
	\bar{Q}_+=\bar{Q}_i\left (1+\xi \right)
	\label{eq2.4.3}
\end{eqnarray}

\noindent Where $\left | \xi \ll1  \right|$. Where $\bar{Q}_i$ is the critical charge. Near the critical point, we perform a Taylor expansion on $\bar{Q}\left ( \bar{r}_+ \right)$ and obtain

\begin{eqnarray}
	\bar{Q}(\bar{r}_+)=\bar{Q}(\bar{r}_i)+\left(\frac{\partial \bar{Q}}{\partial \bar{r}_+}  \right)_c\left(\bar{r}_+-\bar{r}_i \right)+\frac{1}{2}\left(\frac{\partial^2 \bar{Q}}{\partial \bar{r}^2_+} \right)_c\left(\bar{r}_+-\bar{r}_i \right)^2+ \mathcal{O}(\bar{r}_i).
	\label{eq2.4.4}
\end{eqnarray}

\noindent At the critical point where $\left(\frac{\partial \bar{Q}}{\partial \bar{r}_+}  \right)_c =\left(\frac{\partial \bar{Q}}{\partial \bar{T}} \right)_c \left(\frac{\partial \bar{T}}{\partial \bar{r}_+} \right)_c= 0$ , we ignore all higher-order terms $ \mathcal{O}(\bar{r}_i)$, and the above equation can be rewritten as

\begin{eqnarray}
	\bar{Q}(\bar{r}_+)=\bar{Q}(\bar{r}_i)+\frac{1}{2}\left(\frac{\partial^2 \bar{Q}}{\partial \bar{r}^2_+} \right)_c\left(\bar{r}_+-\bar{r}_i \right)^2.
	\label{eq2.4.5}
\end{eqnarray}

\noindent Using Eqs. (\ref{eq2.4.2}), (\ref{eq2.4.3}) and (\ref{eq2.4.5}), we get

\begin{eqnarray}
	\sigma^2= \frac{1}{2} \frac{\bar{Q}_i\xi }{\bar{r}^2_i}\left(\frac{\partial^2 \bar{Q}}{\partial \bar{r}^2_+} \right)_c.
	\label{eq2.4.6}
\end{eqnarray}

\noindent Then we expand the LEs using Taylor series, neglect all higher-order terms and obtain

\begin{eqnarray}
	\lambda(\bar{r}_+)=\lambda(\bar{r}_i)+\left(\frac{\partial \lambda}{\partial \bar{r}_+}\right)_c\left(\bar{r}_+-\bar{r}_i\right).
	\label{eq2.4.7}
\end{eqnarray}

\noindent From Eqs. (\ref{eq2.4.2}), (\ref{eq2.4.3}), (\ref{eq2.4.6}) and (\ref{eq2.4.7}), we get

\begin{eqnarray}
	\lambda(\bar{r}_+)-\lambda(\bar{r}_i)=\left(\frac{\partial \lambda}{\partial \bar{r}_+}\right)_c\left(\frac{1}{2}\frac{\partial^2 \bar{Q}}{\partial \bar{r}^2_+} \right)^{-\frac{1}{2}}_{\bar{r}_+=\bar{r}_i}\left(\bar{Q}-\bar{Q}_i\right)^{\frac{1}{2}}.
	\label{eq2.4.8}
\end{eqnarray}

\noindent Observing the above equation, we find that the critical exponent $\alpha_1$ is equal to $1/2$. Through similar calculations, we obtain

\begin{eqnarray}
	\lambda(\bar{r}_+)-\lambda(\bar{r}_i)=\left(\frac{\partial \lambda}{\partial \bar{r}_+}\right)_c\left(\frac{1}{2}\frac{\partial^2 \bar{T}}{\partial \bar{r}^2_+} \right)^{-\frac{1}{2}}_{\bar{r}_+=\bar{r}_i}\left(\bar{T}-\bar{T}_i\right)^{\frac{1}{2}}.
	\label{eq2.4.9}
\end{eqnarray}

\noindent Therefore, we find that the critical exponents $\alpha_1$ and $\alpha_2$ are both $1/2$. In \cite{GLMW}, $\Delta\lambda=\lambda_s-\lambda_l$ is treated as the order parameter, where $\lambda_s$ and $\lambda_l$ represent the Lyapunov exponents for the small BH and large BH phases, respectively. These studies determined the critical exponents associated with the order parameter to be $1/2$. Clearly, our results are consistent with theirs.

\section{Violation of chaos bound}\label{sec4}

\begin{figure}[h]
	\centering
	\begin{minipage}[t]{0.48\textwidth}
		\centering
		\includegraphics[width=7cm,height=5cm]{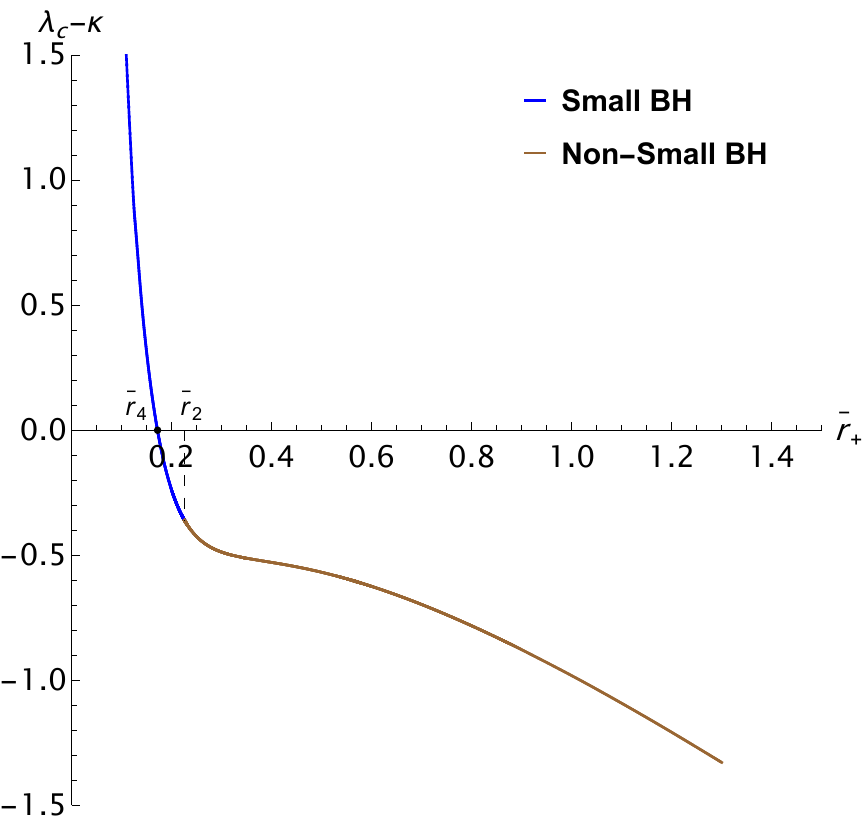}
		\subcaption{}
		\label{4-7-a}
	\end{minipage}
	\begin{minipage}[t]{0.48\textwidth}
		\centering
		\includegraphics[width=7cm,height=5cm]{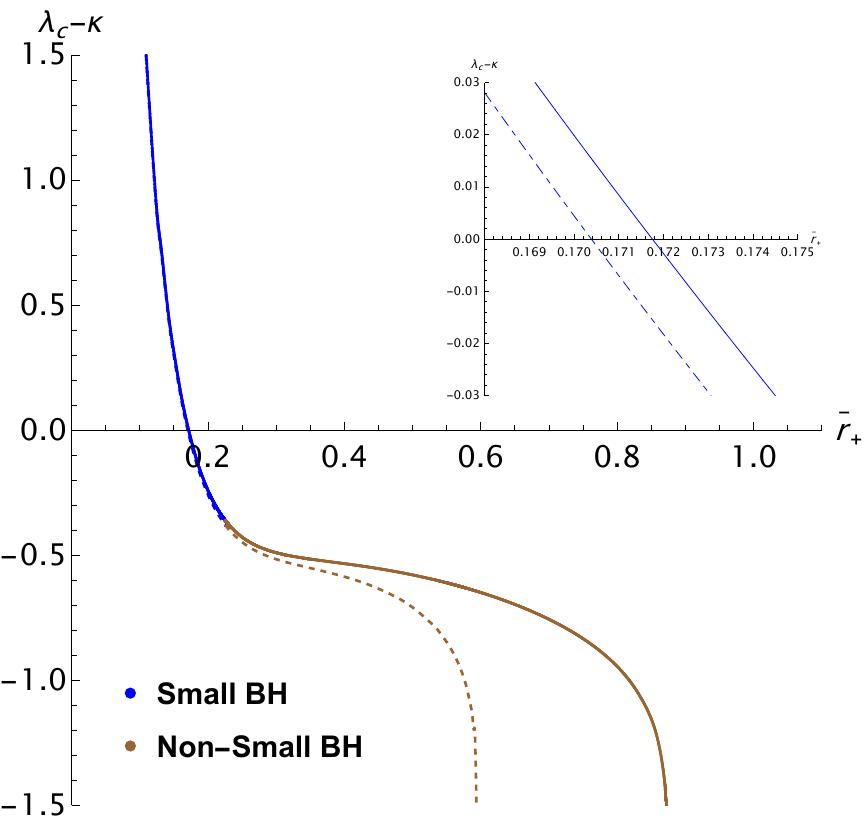}
		\subcaption{}
		\label{4-7-b}
	\end{minipage}
	\caption{ The relationship between the LEs and the horizon radius $\bar{r}_+$ of the Kerr-Newman AdS BH, where $\bar{Q}=0.1\bar{Q}_c$. Figure \ref{4-7-a} illustrates the case of the massless particle. Figure \ref{4-7-b} illustrates the case of the massive particle, where a solid line corresponding to the case of $L=20l$ and a dashed line corresponding to the case of $L=6l$.}
	\label{4f7}
\end{figure}

\begin{figure}[h]
	\centering
	\begin{minipage}[t]{0.48\textwidth}
		\centering
		\includegraphics[width=7cm,height=5cm]{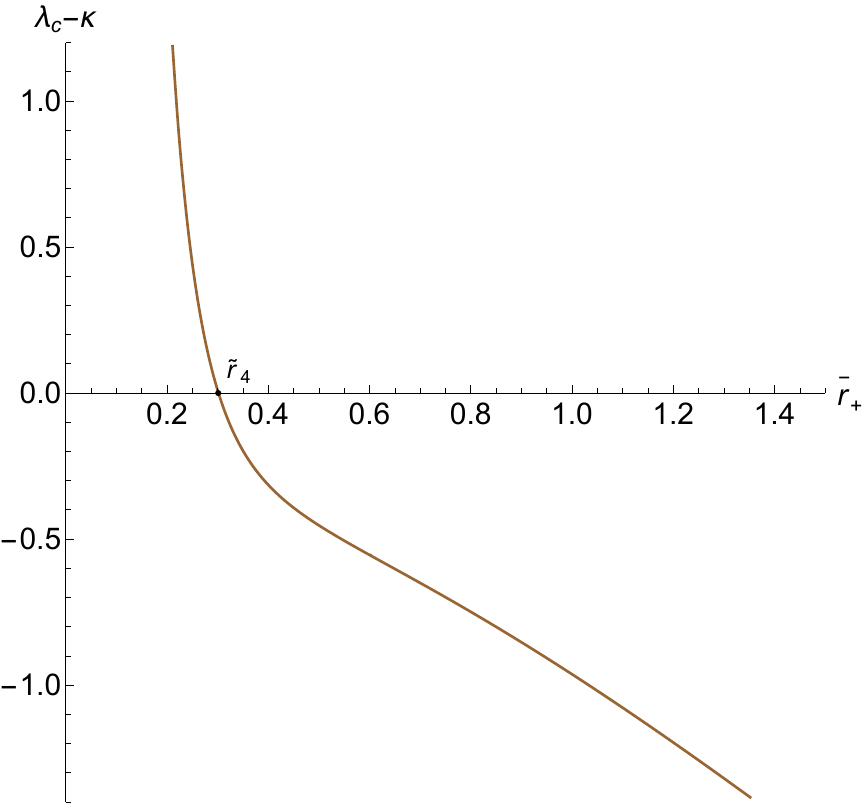}
		\subcaption{}
		\label{4-8-a}
	\end{minipage}
	\begin{minipage}[t]{0.48\textwidth}
		\centering
		\includegraphics[width=7cm,height=5cm]{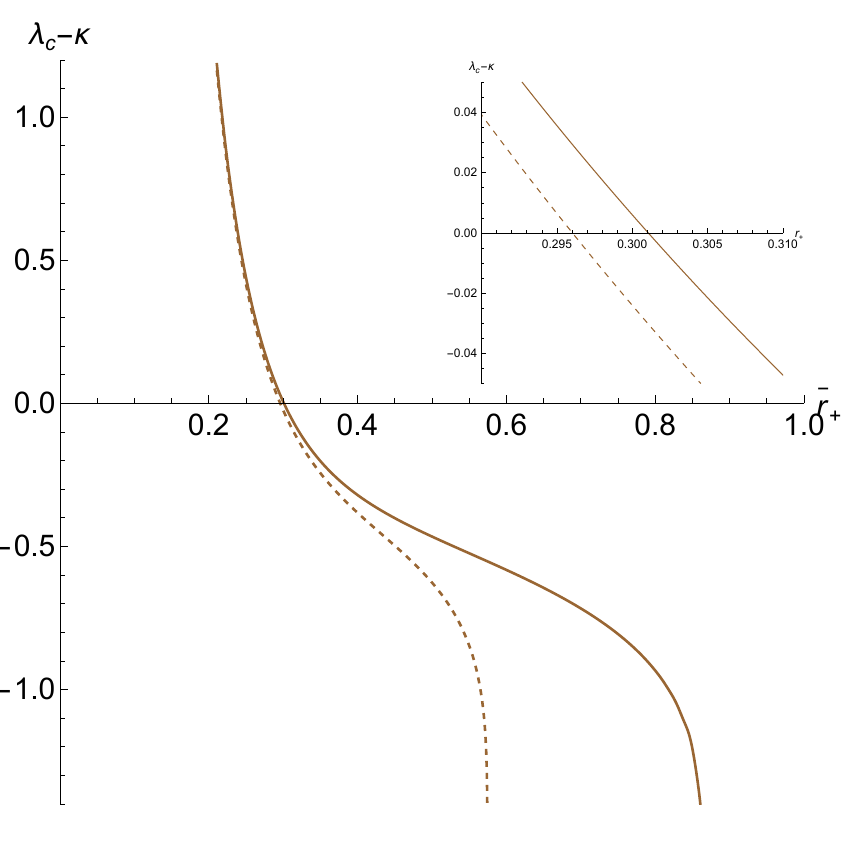}
		\subcaption{}
		\label{4-8-b}
	\end{minipage}
	\caption{ The relationship between the LEs and the horizon radius  $\bar{r}_+$ of the Kerr-Newman AdS BH, where $\bar{Q}=1.5\bar{Q}_c$. Figure \ref{4-8-a} illustrates the case of the massless particle. Figure \ref{4-8-b} illustrates the case of the massive particle, where a solid line corresponding to the case of $L=20l$ and a dashed line corresponding to the case of $L=6l$. }
	\label{4f8}
\end{figure}

In \cite{HT}, Hashimoto and Tanahashi revealed that the upper bound of the LE isn't larger than the surface gravity of the BH, that is $\lambda \le \kappa$, where $\kappa$ denotes the surface gravity. If $\lambda > \kappa$, it indicates that the chaos bound is violated; otherwise, the chaos bound is satisfied. Here, we study the relationship between the upper bound of the LE and the surface gravity of the Kerr-Newman AdS BH during its phase transition. We employ Eqs. (\ref{eq2.2.5}), (\ref{eq2.2.21}),  (\ref{eq2.3.1}), (\ref{eq2.3.2})  and the relation $\kappa=2\pi T$ to plot the difference between the LE and surface gravity against the horizon radius, as shown in Figures \ref{4f7} and \ref{4f8}, respectively.

In Figure \ref{4f7},  the BH charge is fixed at $\bar{Q}=0.1\bar{Q}_c$, the blue curve represents the small BH phase while the brown curve corresponds to other phases. In Figure \ref{4-7-a}, a single solid line appears because the orbital angular momentum of the massless particle does not affect their unstable circular orbits. The radius $\bar{r}_2$ marks the phase transition point where the BH evolves from a small BH phase to a non-small BH phase, corresponding to the temperature $\bar{T}_2$ in Figure \ref{4-2-b}. The curve reveals a violation for the chaos bound when the horizon radius $\bar{r}_+<\bar{r}_4$, with the onset radius $\bar{r}_4$ of the violation satisfying  $\bar{r}_4<\bar{r}_2$. This suggests that the violation of the chaos bound occurs in the region of the stable small black hole. In Figure \ref{4-7-b}, we employ a solid line to plot the relationship between the LE and the horizon radius, where the the orbital angular momentum of the particle is $L=20l$, and a dashed line is the case of $L=6l$. As illustrated in this figure, an increase in the angular momentum $L$ leads to a corresponding expansion of the horizon radius $\bar{r}_+$ at the point where the chaos bound is violated. However, this violation is observed solely during the stable small BH phase, and the radius consistently remains below the threshold radius that distinguishes the small BH phase from the non-small BH phase. The disparity in the exponent's values corresponding to the two distinct angular momenta widens progressively with an increase in the horizon radius. The emergence of this phenomenon is due to that the orbital angular momentum of the massive particle affects their unstable circular orbits, resulting in two distinct curves. 

In Figure \ref{4f8}, the BH's charge is fixed at $\bar{Q}=1.5\bar{Q}_c$, and there is no phase transition. Figure \ref{4-8-a} displays the relationship between the LE for the chaos of the massless particle and the horizon radius $\bar{r}_+$ of the Kerr-Newman AdS BH. In this figure, the violation of chaos bound is observed below the maximum radius $\tilde{r}_4$. By comparing with Figure \ref{4-7-a}, we find that $\tilde{r}_4>\bar{r}_4$, which indicates that an increase of the BH's charge enlarges the critical radius where the chaos bound is violated. Similar to Figure \ref{4-7-b}, we observe in Figure \ref{4-8-b} that as the particle's angular momentum increases, the horizon radius at the critical point where the chaos bound is violated also increases. Consequently, Figures \ref{4f7} and \ref{4f8} collectively suggest that, irrespective of the phase transition occurring within the BH, the violation of the chaos bound takes place when the BH's horizon radius is smaller than the thresholds.

\section{Conclusions and discussions}\label{sec5}

In this paper, we studied the LEs of chaotic motion for both massless and massive particles within the equatorial plane of the Kerr-Newman AdS BH and explored their connection with the BH's phase transition. The results indicated that both the LE of the massless particle and that of the massive particle reveal the phase transition; however, the former is more effective indicator than the latter. To further delve into their connection, we calculated the critical exponents associated with the LEs and found their values to be 1/2.

In a previous work \cite{LGD}, the authors studied the violation of the chaos bound and discovered that such a violation can only occur in the thermodynamically stable phase of the $D$-dimensional RN BHs. Our study further corroborates this finding and demonstrates that when the Kerr-Newman AdS BH undergoes the phase transition, the violation occurs in the small BH's spacetime. Moreover, this violation exists regardless of whether the BH is undergoing a phase transition. When the chaos bound is violated, there exists a critical threshold for the BH. When the horizon radius is smaller than this critical value, the violation of the chaos bound takes place.

\end{document}